\documentclass[aps,preprint]{revtex4}
\usepackage{graphicx}
\usepackage{amsmath}

\newcommand{\eqp}[1]{}

\renewcommand{\cite}{\citep}

\newcommand{\gpfigc}[1]{%
  \resizebox{0.70\textwidth}{!}{\includegraphics{#1}}\\[\baselineskip]
  }

\newcommand\uv{{\bf u}}
\newcommand\cv{{\bf c}}
\newcommand\nv{{\bf n}}
\newcommand\qv{{\bf q}}
\newcommand\dv{{\bf d}}
\newcommand\Rv{{\bf R}}
\newcommand\Vv{{\bf V}}
\newcommand\Hv{{\bf H}}

\newcommand\rv{{\bf r}}
\newcommand\Fv{{\bf F}}
\newcommand\p{{\partial}}
\newcommand\Iv{{\bf I}}

\newcommand\etav{\mbox{\boldmath $\eta$}}

\newcommand\tauroddil{\tau_{rod}^{0}}
\newcommand\zetaperp{\zeta_{\perp}}
\newcommand\zetapar{\zeta_{\parallel}}

\newcommand\msdtubedia{\langle \Delta d^{2} (t) \rangle}
\newcommand\Av{{\bf A}}
\newcommand\zv{{\bf z}}

\begin{document}

\title{A Brownian dynamics algorithm for entangled wormlike threads}

\author{Shriram Ramanathan}
\affiliation{Department of Chemical Engineering and Materials Science,
             University of Minnesota, Minneapolis, MN 55455, USA}
\author{David C. Morse}
\email{morse@cems.umn.edu}
\affiliation{Department of Chemical Engineering and Materials Science,
             University of Minnesota, Minneapolis, MN 55455, USA}
  
\date{\today}

\begin{abstract}
  We present a hybrid Brownian dynamics / Monte Carlo algorithm for 
  simulating solutions of highly entangled semiflexible polymers or 
  filaments.  The algorithm combines a Brownian dynamics 
  time-stepping approach with an efficient scheme for rejecting moves 
  that cause chains to cross or that lead to excluded volume overlaps. 
  The algorithm allows simulation of the limit of infinitely thin 
  but uncrossable threads, and is suitable for simulating the
  conditions obtained in experiments on solutions of long actin
  protein filaments. 
\end{abstract}

  \maketitle

  \section{Introduction}
  \label{sec:intro}
 
  Experimental and theoretical studies of solutions of entangled 
  semiflexible polymers have probed concentration regimes that are
  difficult to simulate by standard methods. In solutions of actin
  protein filaments, the typical filament  length $L$ and persistence
  length $L_{p}$ are both of order  $10\mu$m, but the chain diameter
  $d \simeq 8$ nm is about $10^{3}$  times smaller. For such solutions
  at concentrations near the isotropic-nematic transition, the  actin
  volume fraction is of order $10^{-3}$, $cL^{3} \sim  10^{3}-10^{4}$
  (where $c$ is a number of chains per unit volume), and the typical
  distance between chains is  of order $0.1 \mu m$. These conditions
  may be fruitfully idealized in theoretical work by considering 
  the limit  $d = 0$ of uncrossable but infinitely thin threads.  
  A simulation of these conditions with a conventional molecular
  dynamics (MD) or Brownian dynamics (BD) model of polymers as
  necklaces of nearly-tangent repulsive spherical beads 
  \citep{KremerGrest1990} would require a large number $N \sim L/d 
  \sim 10^{3}$ beads per chain, and an extremely small time step in 
  order to resolve both the strong short range repulsion and the 
  bending forces needed to maintain angular correlations over 1000
  beads. Here, we present a hybrid  Brownian Dynamics (BD) / Monte
  Carlo (MC) algorithm for simulation of such systems of very thin
  threads, in which  chains are represented as sequences of a much
  smaller number of uncrossable rods. The algorithm is designed to
  allow efficient simulation of the idealized limit of infinitely
  thin but uncrossable threads, as well as of chains with a small 
  but nonzero steric diameter.

  The article is organized as follows. Sec. \ref{sec:simalgo}
  is a description of our algorithm for simulating solutions 
  of uncrossable bead-rod chains. Sec. \ref{sec:geometry} is a
  description of the geometrical algorithms used to efficiently
  detect chain crossings and to calculate the distance of closest
  approach between rods. Sec. \ref{sec:validation} presents 
  results of tests of the validity and computational cost of the 
  algorithm. Sec. \ref{sec:analysis} is a discussion of the
  theory underlying the algorithm. Conclusions are summarized 
  in Sec. \ref{sec:conclusions}. 

  \section{Simulation Algorithm}
  \label{sec:simalgo}
  Our simulation method combines a Brownian dynamics time-stepping
  algorithm with a scheme for rejecting moves that cause chains to
  cross or overlap. At each step, a Brownian dynamics (BD) algorithm
  that has been used previously to describe non-interacting wormlike
  chains is used to generate a trial move for a randomly chosen 
  chain. To simulate a solution of infinitely thin but uncrossable 
  threads, we reject all trial moves that cause one chain to cross 
  through another, and accept all others. 

  \subsection{Single Chain Brownian Dynamics}
  \label{sub:Brownian}
  The simulations use a discretized model of wormlike chains in 
  which each chain in a solution is represented as a set of $N+1$ 
  beads, which act as point sources of friction, connected by $N$ 
  inextensible rods, each of which is constrained to have a fixed
  length $a$. Let $\Rv_{\mu}$ be the position of bead $\mu$ of a 
  chain, with $\mu=1,\ldots,N+1$, and $\qv_{i} = \Rv_{i+1}-\Rv_{i}$ 
  be the vector associated with rod $i$, which is constrained to
  have a fixed length $|\qv_{i}| = a$, for all $i=1,\ldots,N$. Let 
  $\uv_{i} = \qv_{i}/|\qv_{i}|$ be a unit tangent vector for rod
  $i$. The bending energy, denoted by $U_{0}$, is 
  \begin{equation}
     U_{0} = -  \frac{\kappa}{a}
     \sum_{i=2}^{N} \uv_{i} \cdot \uv_{i-1}
     \quad,
  \end{equation}
  where $\kappa$ is the bending rigidity of the chain. The
  persistence length is $L_{p} = \kappa /k_{B}T$. This model
  approximates the bending energy of a continuous wormlike 
  chain in the limit $a \ll L_{p}$ and $N \gg 1$. 

  To generate trial moves for individual chains, we use a Brownian 
  dynamics algorithm for chains with constrained rod lengths that 
  has used previously to study dilute solutions of both semiflexible
  \citep{everaersetal1999,pasquali2001,dimitrakapolousetal2001,Shankar2002,matteo2002,pasquali2005} 
  and flexible chains \cite{hinch1994}. In this algorithm, changes in 
  bead positions are calculated from a bead velocity $\Vv_{\mu}$
  of the form
  \begin{equation}
    \label{eq:simul-velocity-semiflex}
    \Vv_{\mu} = \Hv_{\mu} \cdot \left [ -\frac{\p U_{0}}{\p \Rv_{\mu}} 
    + \mathcal{T}_{\mu} \uv_{\mu} - \mathcal{T}_{\mu-1} \uv_{\mu-1} 
    + \Fv_{\mu}^{met} + \etav_{\mu} \right ].
  \end{equation}
  Here, $\Hv_{\mu}$ is a mobility tensor for bead $\mu$, $\mathcal{T}_{i}$ 
  is a tension in rod $i$, and $\etav_{\mu}$ is a random Langevin force. 
  $\Fv_{\mu}^{met}$ is a metric correction force that is required in 
  order for this algorithm to yield the correct equilibrium distribution 
  \cite{fixman1978,hinch1994,matteo2002,Morse2004,pasquali2005}.  

  We use an anisotropic bead mobility tensor $\Hv_{\mu}$ of the form 
  \cite{pasquali2005}
  \begin{equation}
     \Hv_{\mu} = 
     \frac{a}{\zetapar}\tilde{\uv}_{\mu}\tilde{\uv}_{\mu} +
     \frac{a}{\zetaperp}(\Iv - \tilde{\uv}_{\mu}\tilde{\uv}_{\mu})
     \quad,
  \end{equation}
  where $\tilde{\uv}_{\mu}$ is an approximation to the unit  tangent
  vector at bead $i$, and $\zetapar$ and $\zetaperp$ are friction
  coefficients per unit length for parallel and perpendicular  motion,
  respectively. \cite{pasquali2005}. The unit tangent
  $\tilde{\uv}_{\mu}$ is approximated by a centered difference
  $\tilde{\uv}_{\mu} = (\Rv_{\mu+1}-\Rv_{\mu-1})/
  |\Rv_{\mu+1}-\Rv_{\mu-1}|$ for beads $\mu=2,\ldots,N$, and by
  $\tilde{\uv}_{1}=\uv_{1}$ and $\tilde{\uv}_{N+1}=\uv_{N}$ for  the
  end beads.

  The random force $\etav_{\mu}$ is generated by the procedure
  described by \citet{Morse2004} and \citet{pasquali2005} for 
  generating ``geometrically projected'' random  forces. At the 
  beginning of each  time step, an unprojected random force 
  $\etav_{\mu}'$ for each bead is generated from a distribution
  with a vanishing mean, $\langle \etav_{\mu}' \rangle = 0$ and 
  a variance $\langle \etav_{\mu}' \etav_{\nu}' 
  \rangle = 2 k_{B} T  \Hv^{-1}_{\mu} \delta_{\mu\nu}/\Delta t$.
  The force $\etav_{\mu}$ in Eq. 
  (\ref{eq:simul-velocity-semiflex}) is a projected force of 
  the form $\etav_{\mu} = \etav_{\mu}' - \hat{\eta}_{\mu}
  \uv_{\mu} - \hat{\eta}_{\mu-1}\uv_{\mu-1}$ in which the 
  quantities $\hat{\eta}_{1},\ldots,\hat{\eta}_{N}$ are 
  calculated by requiring that 
  $0 = \uv_{i} \cdot ( \etav_{i+1} - \etav_{i} )$ 
  for all $i=1,\ldots,N$. 

  The tensions $\mathcal{T}_{1},\ldots,\mathcal{T}_{N}$ are 
  calculated, after calculation of the bending, metric, and random 
  forces, by requiring that the rod lengths all maintain constant 
  length, or that $0 = \uv_{i} \cdot ( \Vv_{i+1} - \Vv_{i} )$ 
  for all $i=1,\ldots,N$. This yields a tridiagonal set of $N$ 
  equations, which must be solved every time step. 

  Each time step for a single chain is generated by a mid-step 
  integration algorithm originally proposed by Fixman 
  \cite{fixman1978,Morse2004}. At the beginning of each time
  step, projected random forces are generated. Predicted 
  mid-step bead positions are then calculated as
  \begin{equation}
     \Rv_{\mu}^{(1/2)} = \Rv_{\mu}^{(0)} + \Vv_{\mu}^{(0)} \Delta t / 2 
  \end{equation}
  where $\Rv_{\mu}^{(0)}$ represents an initial bead position
  and $\Vv_{\mu}^{(0)}$ represent a bead velocity calculated 
  using the initial bead positions. Final bead positions are 
  calculated as
  \begin{equation}
     \Rv_{\mu}^{(1)} = \Rv_{\mu}^{(0)} + \Vv_{\mu}^{(1/2)} \Delta t
  \end{equation}
  where $\Vv_{\mu}^{(1/2)}$ is a bead velocity computed using 
  bending and metric forces, mobilities, and tensions that are
  re-calculated using the mid-step bead positions, but using 
  the same random force $\etav_{\mu}$ as that used to calculate 
  the mid-step positions.

  \subsection{Detecting chain crossing}
  \label{sec:bdalgo-entangled-semiflex-thin}
  To simulate a solution of infinitely thin but uncrossable 
  threads, we combine this Brownian dynamics algorithm with 
  a scheme for rejecting moves that cause chains to cross. 
  At every timestep, a trial move is generated for a randomly 
  chosen polymer chain using the Brownian dynamics algorithm 
  discussed above.  The trial move is rejected, and the chain 
  position is unchanged, if the trial move would cause any of 
  the N rods of the chosen chain  to cut through any rod of 
  any other chain. Whether or not the move is accepted,
  another chain is then chosen at random, and the process is
  repeated. To accurately simulate Brownian motion, the time 
  step $\Delta t$ used in the Brownian dynamics algorithm must 
  be chosen to be small enough so that almost all of the moves 
  are accepted. In a system containing $M$ chains, a time 
  $\Delta t$ is taken to have elapsed every $M$ attempted chain 
  moves.

\begin{figure}
    \begin{center}
      \gpfigc{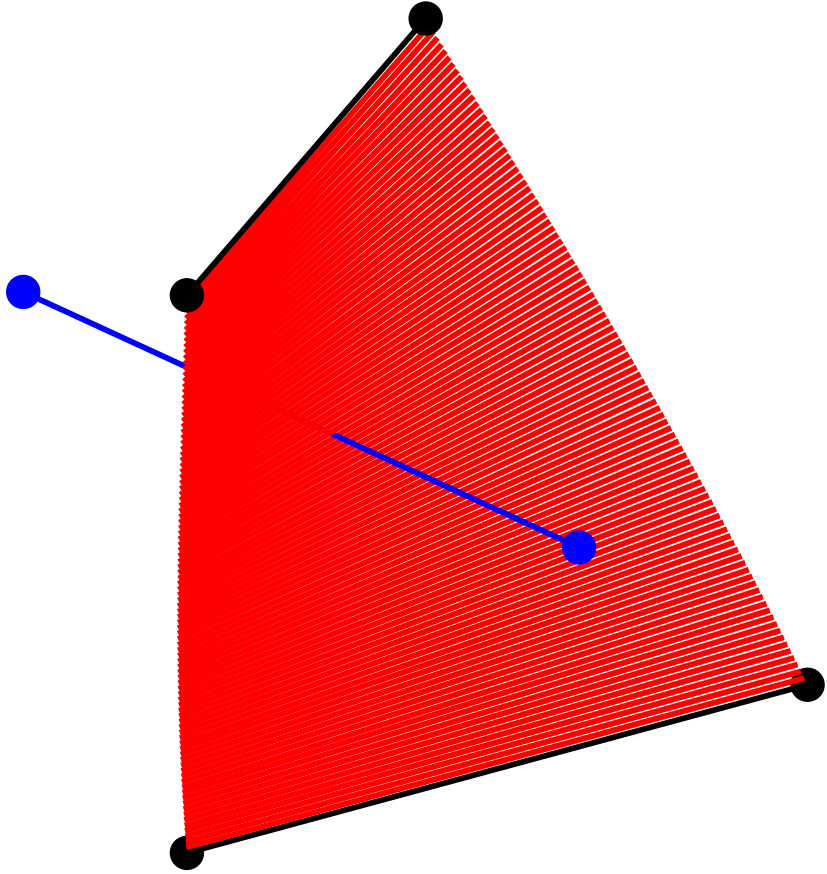}
      \caption[Algorithm for checking topological
      violations]{Intersection of a stationary rod with the surface
      swept out by a moving rod}
      \label{fig:rod-plane-intersect}
    \end{center}
  \end{figure}

  The usefulness of the algorithm relies upon the availability 
  of an efficient method for detecting when a trial move causes 
  one chain to cut through another. Consider an attempted move 
  of a chain $\alpha$. To detect whether this attempted move 
  would cause chain $\alpha$ to cut through any other chain, we 
  consider each of the $N$ rods of chain $\alpha$ to sweep out
  a surface over the course of a time step. We reject the move
  of the chain if the surface swept out by any rod of $\alpha$ 
  is intersected by any rod of any other polymer $\beta$, as
  show schematically in Figure \ref{fig:rod-plane-intersect}.

  Consider the test for whether a move of rod $i$ of the moving
  chain $\alpha$ will cause it to cut through a rod $j$ of a 
  stationary chain $\beta$. The line segment corresponding to 
  the stationary rod (rod $j$ of $\beta$) may be described by 
  a parametric equation
  \begin{equation}
    \label{eq:rod-parametric-eqn}
    \rv'(s') = \cv' + s' \qv'.
  \end{equation}
  where $\cv'$ is the center of the rod, $\qv'$ is the vector 
  connecting its ends, which is of length $|\qv'|=a$, and $s'$ 
  is a contour length parameter with a domain $-1/2 \leq s' 
  \leq 1/2$.  The surface swept out by the moving rod (rod 
  $i$ of $\alpha$) during a trial move can be represented by 
  a two parameter equation,
  \begin{equation}
      \label{eq:rodsurface}
      \rv (s,t) = \cv(t) + s \qv(t),
  \end{equation}
  where $-1/2 < s < 1/2$, $t$ is a dimensionless time variable 
  that ranges from $0$ to $1$ during the time step, and $\cv(t)$ 
  and $\qv(t)$ are the center position and end-to-end vector of 
  the rod as functions of $t$.  Thus, $\rv(s,0)$ and $\rv(s,1)$ 
  represent rod conformations at the beginning and end of the 
  timestep, respectively. To define the surface swept out at
  intermediate times, we take $\cv(t)$ and $\qv(t)$ to be 
  linear functions of $t$, 
  \begin{eqnarray}
    \label{eq:com-orient-time}
    \rv(t) &=& \cv_{0} + t \; \Delta\cv \nonumber \\
    \qv(t) &=& \qv_{0} + t \; \Delta \qv
  \end{eqnarray}
  where $\cv_{0} = \cv(0)$, $\qv_{0}=\qv(0)$, 
  $\Delta\cv \equiv \cv(1) - \cv(0)$, and 
  $\Delta \qv \equiv \qv(1) - \qv(0)$. This is equivalent
  to assuming that the beads at both ends of the rod move 
  with constant velocity from initial to final position, 
  and that the rod is always a straight line between these 
  two beads. The length of the rod is not exactly conserved
  at intermediate times, but the surfaces swept out by
  neighboring rods within a chain meet without gaps or
  overlaps along the trajectory of the shared bead.  

  For the surface $\rv(s,t)$ swept out by the moving rod to
  be intersected by a stationary rod $\rv'(s')$, the line
  segments $\rv'(s')$ and $\rv(s,t)$ must intersect at some 
  time $t=t_{I}$, such that $0 < t_{I} < 1$, at a point of 
  intersection for which $-1/2<s<1/2$ and $-1/2<s'<1/2$. 
  An efficient algorithm for determining whether such an
  intersection occurs for any specified pair of rods is 
  described in Sec. \ref{sec:geometry}.

  \subsection{Rod-Rod Interactions}
  \label{sec:bdalgo-entangled-semiflex-fat}
  The algorithm for infinitely thin threads can be generalized to
  allow for intermolecular interactions. We consider a model in 
  which the total potential energy is a sum of a bending energy
  $U_{0}$ and an interaction potential $U_{int}$. The interaction
  energy is taken to be a pairwise additive sum of interactions 
  between rods,
  \begin{equation}
    \label{eq:U-int-soft-int}
    U_{int} = \sum_{i > j} U ( | \dv_{ij} | ),
  \end{equation}
  where the sum is taken over all pairs of rods, and where the
  interaction energy $U(|\dv_{ij}|)$ between rods $i$ and $j$ is  a
  function of the distance of closest approach $\dv_{ij}$ between
  those two rods.  An algorithm for calculating distance of  closest
  approach between two finite line segments, and a discussion of  how
  this distance is defined, is given in Sec. \ref{sec:geometry}.

  In an algorithm in which trial moves are generated using a
  Brownian dynamics algorithm for non-interacting chains with
  potential energy $U_{0}$, trial moves that do not cause
  chains to cross are accepted or rejected according to a 
  Metropolis acceptance criterion that is based on the change
  in interaction energy: A trial move is accepted if it causes 
  $U_{int}$ to decrease, and is accepted with a probability
  \begin{equation}
     P_{acc} = \exp [- \Delta U_{int} / (k_{B} T) ] 
     \label{PaccUint}
  \end{equation}
  if it leads to an increase $\Delta U_{int} > 0$ in $U_{int}$.
  In the case of a hard-core interaction, a move that does 
  not cause chain crossing is rejected if it causes the distance
  $|\dv_{ij}|$ between any two rods to fall below a specified 
  hard-core diameter, and accepted otherwise. Note that only
  the interaction energy is used in this acceptance rule,
  and not the change in the intramolecular bending potential, 
  because the effect of the bending potential is already 
  fully accounted for by the Brownian dynamics algorithm used 
  to generate trial moves. The rationale underlying the use of 
  such a Metropolis sampling is discussed in Sec.
  \ref{sec:analysis}.

  Each step of the simulation thus involves the following
  sequence of operations. A trial move is generated for a
  randomly chosen chain. A check is performed for every 
  rod on that chain to see if the trial move would cause
  intersection of that rod with any of the neighboring
  rods, which are identified by use of a Verlet neighbor 
  list. The move is rejected if any intersections are 
  found. If no intersections are found, and $U_{int} \neq 0$,
  the change in the interaction energy of each rod with 
  all nearby neighbors is calculated, and the move 
  is accepted or rejected based upon the Metropolis 
  criterion described above.
  
  As in simulations of systems of point particles
  \citep{allentildesley}, Verlet neighbor lists are  used to reduce
  the number of rods that must be  checked for intersections and for
  which interaction  energy must be calculated.  The neighbor list for
  each  rod in the system contains all other rods for which the
  distance of closest approach was less than a  Verlet radius $r_{l}$
  when the neighbor list was constructed.  The neighbor list is
  reconstructed whenever any bead  in the simulation is found to have
  moved a distance  greater than  $r_{l}/2$ since the last update of
  the  list. The neighbor list is constructed with the use of a linked
  cell list of rods, by a procedure closely  analogous to that used for
  point particles. Optimal values of the Verlet radius for each set of
  parameters have been determined by extensive trial and error.

  \subsection{Preparation of initial states}
  \label{sec:InitialConfig}
  Many of the simulations for which we have used our algorithm
  focus on the dynamics and stress relaxation in entangled 
  solutions over relatively short time scales. These simulations
  are often run for times shorter than a reptation time due to 
  computational limitations on our ability to simulate very 
  crowded solutions for long times. For such short simulations 
  to yield valid results, they must start from initial 
  configurations that  are already representative of a thermal 
  equilibrium ensemble for the solution. Such simulations thus 
  rely upon our ability to efficiently generate equilibrated 
  initial states for systems that would be difficult or 
  impossible to equilibrate by running the algorithm described
  above, which we use to study dynamical properties. 

  In the case of infinitely thin but uncrossable chains, with 
  $U_{int}=0$, the prohibition on chain crossing has an 
  enormous effect upon dynamics of the system, but has no 
  effect upon the equilibrium probability distribution, since 
  it changes neither the set of allowed microstates nor the 
  energy of any microstate.  The equilibrium distribution for 
  a solution of such uncrossable chains is thus identical to 
  that of a solution containing an equal number of non-interacting
  ``phantom'' chains.  This distribution is characterized by 
  random chain positions and orientations, and by a Boltzmann 
  distribution 
  \begin{equation}
    \label{eq:angulardist}
    P(\cos \, \theta_{ij}) = 
    \frac{ e^{(L_{p}/a) \cos\theta_{ij} } }
    { \int_{0}^{\pi} d\theta_{ij} \sin \theta_{ij} 
      e^{(L_{p}/a)\cos\theta_{ij}} },
  \end{equation}
  for the cosine $\cos\theta_{ij} = \uv_{i} \cdot\uv_{j}$ 
  of the angle between neighboring rods $j=i \pm 1$ within 
  a chain.  We may thus generate initial configurations for 
  such a solution by a method used previously to generate 
  initial configurations for dilute solutions, in which 
  chains are grown by a process that yields an equilibrium
  distribution of chain conformations. The first bead of 
  each chain is placed at random within the simulation cell,
  and the first rod is given a random orientation. The 
  orientation of each subsequent rod relative to the previous
  one is chosen such that the cosine of the angle between
  these two rods is chosen from the Boltzmann distribution 
  given in equation \eqref{eq:angulardist}.  

  When $U_{int} \neq 0$, inter-molecular interactions lead to
  non-trivial correlations. In this case, we start with an
  initial configuration for a non-interacting system, generated
  by the method described above, and then equilibrate the 
  system by a slithering-snake Monte-Carlo algorithm. In each 
  attempted move of this algorithm, a rod is removed from 
  one end (the ``tail'') of a randomly chosen polymer and reattached 
  to the other end (the ``head''). The orientation of the new rod
  relative to that of the pre-existing end rod is chosen from  the
  equilibrium distribution given in \eqref{eq:angulardist}), where 
  $\cos\theta$ is the angle of the last joint in the chain. The
  resulting attempted Monte-Carlo move is accepted or rejected
  based on a Metropolis criterion based on the resulting change
  in interaction energy $U_{int}$, as given in Eq. (\ref{PaccUint}) 
  for the acceptance probability of a move that changes the 
  interactions by an amount $U_{int} > 0$.  Monte Carlo equilibration 
  is continued until every chain in the simulation box has reptated 
  many times its own length. In the concentration range of interest, 
  where the system is highly entangled but semi-dilute, and in which 
  the equilibrium phase is isotropic rather than a nematic liquid 
  crystal, almost all such Monte Carlo moves are accepted, and the 
  system can be equilibrated quite rapidly. 

  \section{Geometrical Algorithms}
  \label{sec:geometry}

  \subsection{Intersection}
  To determine whether two line segments intersect during 
  a move, we first determine whether the two infinite lines 
  containing these segments intersect during the time $0 < 
  t < 1$ of interest.  If an intersection time $0 < t_{I} 
  < 1$ is found, we then calculate the contour variables
  $s$ and $s'$ at the point of intersection, and determine
  if these lie within the two line segments that define 
  the rod.

  The efficiency of our algorithm relies upon the existence 
  of efficient algorithms for detecting intersections 
  between rods and calculating distances of closest approach 
  between line segments.

  An intersection between a moving line $\rv(s,t)$ and a 
  stationary non-parallel line $\rv'(s')$ that pass through 
  rod centers $\cv(t)$ and $\cv'$, respectively, can occur at 
  a time $t_{I}$ if the vector $\cv(t_{I}) - \cv$ has a 
  vanishing projection onto a vector $\qv(t_{I}) \times \qv'$ 
  that is perpendicular to both lines, i.e., if
  \begin{equation}
     \label{eq:numereqzero}
     0 = [ \, \cv(t_{I}) - \cv' \, ] \cdot 
         [ \, \qv(t_{I}) \times \qv' \, ]
  \end{equation}
  Substituting Eq. (\ref{eq:com-orient-time}) for $\cv(t) $ and 
  $\qv(t)$ into this condition yields a quadratic equation
  \begin{equation}
    a t_{I}^{2} + b t_{I} + c = 0 \quad .
    \label{eq:quadraticeqn}
  \end{equation}
  for the intersection time $t_{I}$ of the two infinite lines, in 
  which
  \begin{eqnarray}
     \label{eq:coeffs}
     a &=& \Delta\cv \cdot ( \Delta \qv \times \qv' )
     \nonumber \\ 
     b &=& \Delta \cv \cdot ( \qv_{0} \times \qv' )
        + [ \, \cv_{0} - \cv' \, ] \cdot ( \Delta \qv \times \qv' )
     \nonumber \\
     c &=& [ \, \cv_{0} - \cv' \, ] \cdot ( \qv_{0} \times \qv' )
  \end{eqnarray}
  If a real solution is not found in the domain $0<t_{I}<1$, 
  then these two lines do not intersect during the time 
  step of interest. 

  If the infinite lines containing the rods do intersect 
  at a time $0 < t_{I} < 1$, we must then calculate the 
  coordinates $s$ and $s'$ at the point of intersection,
  where $\rv(s,t_{I}) = \rv(s')$. The two rods intersect
  only if these coordinates satisfy $-1/2 < s < 1/2$ and 
  $-1/2 < s' < 1/2$.  We calculate the coordinates $s$
  and $s'$ as a special case of the algorithm for 
  determining the point of closest approach of two lines,
  which is discussed below. 

  Ideally, these geometrical checks should be performed 
  for every rod of the moving chain $\alpha$ and every other 
  rod in the simulation box, including other rods on chain 
  $\alpha$.  The algorithm described above,  however, is 
  designed to check only for intersections of rods on a 
  moving chain with rods on a different stationary chain.
  It  is possible to generalize the algorithm so as to 
  identify intersections between two moving rods. The 
  generalization requires that the quadratic equation for 
  $t_{I}$, equation \eqref{eq:quadraticeqn}, be replaced by a 
  cubic equation.  However, intersections between different 
  parts of the same chain are expected to be rare in the
  systems of rodlike chains upon which we have focused and 
  so, for simplicity, we have not implemented a check for 
  such self-intersections.

  \subsection{Distance of Closest Approach}
  Consider two rods that correspond to line segments
  $-1/2 \leq s \leq 1/2$ and $-1/2 \leq s \leq 1/2$ 
  of the lines
  \begin{eqnarray}
     \rv(s)   & = & \cv + s \qv \nonumber \\
     \rv'(s') & = & \cv + s' \qv' \quad.
     \label{rrofs}
  \end{eqnarray}
  The distance of closest approach (DCA) between these 
  two rods is the minimum magnitude $|\dv(s,s')|$ of 
  the separation
   \begin{equation}
       \dv(s,s') = \rv(s) - \rv(s')
       \label{dvdef}
   \end{equation}
   for $-1/2 \leq s \leq 1/2$ and $-1/2 \leq s' \leq 1/2$.
   This minimum may be obtained when $s$ and/or $s'$ are
   equal to $\pm 1/2$, in which case the distance of 
   closest approach is the distance between the end of
   one rod and a point along the length of the other, 
   or the distance between two rod ends. 

   The first step in the calculation of this minimum
   distance is the identification of the coordinates 
   $s$ and $s'$ at the point of closest approach of 
   the infinite lines that contain the two rods. To 
   identify this point of closest approach, we require 
   that
   \begin{eqnarray}
      0 & = & \partial |\dv|^{2}/\partial s
        = 2 \qv\cdot\dv(s,s') \nonumber \\
      0 & = & \partial |\dv|^{2}/\partial s'
        = 2 \qv'\cdot\dv(s,s') 
   \end{eqnarray}
   Note that this is equivalent to the geometrical 
   requirement that the vector $\dv(s,s')$ at the point 
   of closest approach be perpendicular to both $\qv$ 
   and $\qv'$. Substituting Eqs. (\ref{rrofs}) and 
   (\ref{dvdef}) into these conditions yields a pair of 
   linear equations
   \begin{eqnarray}
     a s - bs' & = & d 
     \label{eq:pca-L1}
     \\
    -b s + cs' & = & e 
     \label{eq:pca-L2}
   \end{eqnarray}
   in which $a = |\qv|^{2}$, $b =\qv\cdot\qv'$, $c=|\qv'|^{2}$, 
   $d =\qv\cdot[\cv' - \cv]$,  and $e=\qv'\cdot[\cv - \cv']$,
   which have the solution
   \begin{eqnarray}
     \label{eq:pca-L1L2-solns}
     s  &=& \frac{be+cd}{ac-b^{2}} \nonumber \\
     s' &=& \frac{ae+bd}{ac-b^{2}} \quad.
   \end{eqnarray}
   As part of the check for the intersection of two rods,
   we use Eq. (\ref{eq:pca-L1L2-solns}) to calculate the 
   coordinates $s$ and $s'$ for pairs of lines $\rv(s) = 
   \rv(s,t_{I})$ and $\rv'(s')$ that have already been 
   shown to intersect at a time $t_{I}$, for $0<t_{I}<1$. 
   In this case the point of closest approach is actually 
   the point of intersection. 

   The distance of closest approach between the non-parallel 
   rods is the same as that for the corresponding infinite 
   lines if and only if the coordinates $s$ and $s'$ at the 
   point of closest approach of the lines satisfy the 
   inequalities $-1/2 < s,s' < 1/2$.  Otherwise, the closest 
   approach between rods may correspond to the separation 
   between an end of one rod and a point along the length 
   of the other (e.g., $s = 1/2$ and $-1/2 < s' < 1/2$), or 
   between two rod ends (e.g., $s= 1/2$ and $s' = -1/2$). To 
   compute the shortest distance between rods, we first use 
   Eq. \eqref{eq:pca-L1L2-solns} to calculate the coordinates 
   $s$ and $s'$ of the point of closest approach for the 
   infinite lines. If both $s$ and $s'$ lie between $-1/2$ 
   and $1/2$, we accept these values and calculate the 
   distance $|\dv(s,s')|$. If not, we must consider one of 
   two possible types of special cases.

   The first special case occurs when one of the two 
   coordinates $s$ and $s'$ at the point of closest approach 
   of two infinite lines lies within the range [-1/2,1/2], 
   and the other does not. As an example, consider the case 
   $s > 1/2$ and $-1/2 \leq s' \leq 1/2$. In this example, 
   the distance of closest approach will be between the rod 
   end $s = 1/2$ and some point $-1/2 \leq s' \leq 1/2$. In 
   this example, we may identify the minimum $|\dv(1/2,s')|$ 
   for unrestricted $s'$ by solving linear equation 
   (\ref{eq:pca-L2}) for $s'$ while taking $s=1/2$.  If the
   resulting value of $s'$ lies outside $[-1/2,1/2]$, then
   the DCA between rods is obtained by setting $s'$ to the 
   value $s = \pm 1/2$ of the nearest end, giving a DCA
   equal to the distance $|\dv(1/2,\pm 1/2)|$ between two 
   rod ends. 

   The other type of special case occurs when neither of 
   the coordinates at the point of closest approach of the 
   infinite lines lies within [-1/2,1/2]. As an example, we 
   consider $s > 1/2$ and $s' > 1/2$. In this case we must 
   locate the minima of both $|\dv(1/2,s')|^{2}$ for 
   $-1/2 \leq s \leq 1/2$ and of $|\dv(s,1/2)|^{2}$ for 
   $-1/2 \leq s \leq 1/2$ by the procedure outlined above, 
   either or both of which may yield $s=s'=1/2$, and accept 
   whichever yields a smaller distance. 

   The above reasoning applies only to non-parallel rods. 
   Two rods are parallel if and only if the denominator 
   in Eq. \ref{eq:pca-L1L2-solns} vanishes, i.e., if 
   $ac-b^{2} = 0$. For parallel rods, if $|d| \leq 
   |a+b|/2$, then the DCA is the magnitude of the projection 
   of $\cv - \cv'$ onto the plane perpendicular to $\qv$, 
   which is given by $|\cv'-\cv -\qv d/a|$.  If $|d| > 
   |a+b|/2$, then the DCA is the distance between two rod 
   ends. 
   
  \section{Validation and results}
  \label{sec:validation}

  \subsection{Equilibrium Properties}
  \label{sec:results-eqblm-properties}
  To show that our algorithm simulates equilibrated solutions  of
  infinitely thin chains, we begin by analyzing results for two
  equilibrium properties for which analytical predictions are
  available.

\begin{figure}
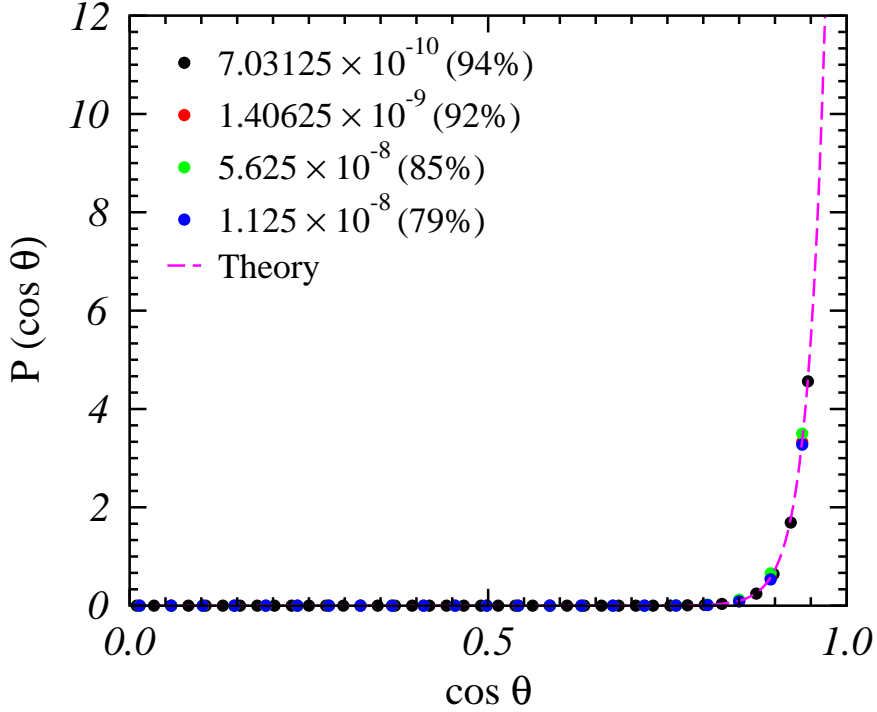

    \begin{center}
    \gpfigc{figs/costheta.eps}
    \caption[Probability distribution for angle between neighboring
     rods on the same chain]{Probability distribution for cosine of
     the angle between neighboring rods of the same chain.
     $cL^{3}=4000$, $N=40$, $L_{p}=L$,
     $\zetapar=\zetaperp$. Legends indicate $ \Delta t $ in units
     of $\tauroddil$.  Values in parenthesis represent percentage
     of trial moves accepted.}
    \label{fig:costheta}
    \end{center}
  \end{figure}

  Figure \ref{fig:costheta} shows simulation results for a histogram 
  of the distribution of angles between neighbouring rods. The angular 
  distributions obtained for dilute and concentrated solutions agree 
  with each other and with equation \eqref{eq:angulardist} within our 
  statistical error. 

\begin{figure}
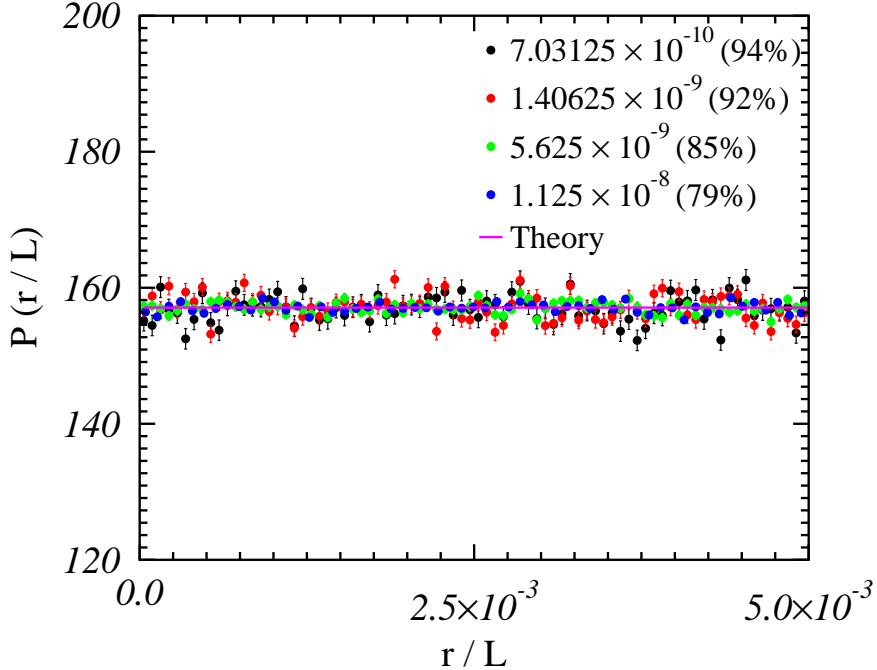

    \begin{center}
      \gpfigc{figs/radial-dist.eps}
      \caption[Radial distribution for rods used to discretize the
      chains]{Radial distribution function for rods used to discretize
        polymer chains. $ cL^{3}=4000 $, $ N=40 $, $ L_{p}=L$,
        $\zetapar=\zetaperp$. Legends indicate $ \Delta t $ in units
        of $\tauroddil$. Values in parentheses represent percentage of
        trial moves accepted.}
      \label{fig:raddist}
    \end{center}
  \end{figure}

  Figure \ref{fig:raddist} plots a radial distribution function
  obtained from simulations of entangled semiflexible polymers.  
  The radial distribution function $P(r)$ for a test rod $i$ 
  is defined here such that $P(r)dr$ is the probability per unit 
  length of the test rod of finding another rod $j$ for which 
  the points of closest approach of the lines containing the 
  two rods lies within the line segments represented by the 
  rods, and for which the distance of closest approach lies 
  between $r$ and $r+dr$. We show in appendix \ref{app:radial-dist-fn} 
  that $P(r)$ for a solution of infinitely thin, randomly 
  distributed chains is given by
  \begin{equation}
    P(r) = \frac {\pi} {2} \rho
    \label{eq:raddistfn}
  \end{equation}
  The radial distribution functions obtained from our simulations 
  agree with that predicted by equation \eqref{eq:raddistfn} to 
  within statistical error. This agreement extends to very small 
  values of $r$, comparable to the distance moved by the rod 
  within a single timestep $\Delta t$, for which we might have 
  expected our rejection scheme for preventing chain crossing 
  to distort the equilibrium distribution.
  
  \subsection{Convergence studies}
  \label{sec:convergence-deltat-nrods}
  Results presented in Sec. \ref{sec:results-eqblm-properties}
  clearly show that statistical equilibrium properties are well 
  converged with respect to the integration timestep $\Delta t$.  
  However, this does not necessarily imply that dynamic properties 
  are simulated with equal accuracy. 

\begin{figure}
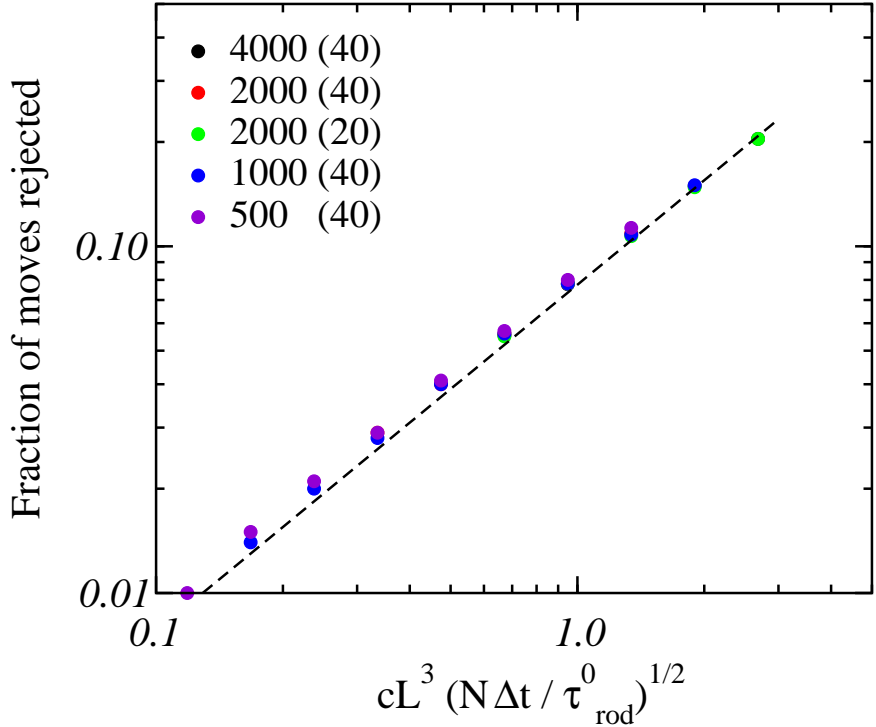

    \begin{center}
      \gpfigc{figs/percent-reject.eps}
      \caption[Fraction of trial moves rejected as a function 
      $ \Delta t $]{Fraction of trial moves rejected as a 
      function of $cL^{3}\sqrt{N\Delta t/\tau_{rod}^{0}}$.
      Here, $\tau_{rod}^{0} = \zeta_{\perp}L^{3}/(72k T)$,
      $N$ is the number of rods per chain. Numbers in the 
      legend correspond to values of $cL^{3}$ and (in 
      parenthesis) of $N$. All simulations are for $L=L_{p}$. 
      Dashed line represents $y = 0.0775 x$}
      \label{fig:percentreject}
    \end{center}
  \end{figure}

  We expect dynamical properties to be sensitive to the fraction 
  of trial moves that are rejected, which increases with increasing
  time step $\Delta t$. The dependence of the fraction of moves
  rejected on $\Delta t$ and other parameters can be understood
  by the following scaling argument: The probability $P_{reject}$
  that a chain of $N$ rods and length $L=Na$ will intersect another 
  chain during a trial move is roughly $P_{reject} \sim P(r) N
  a \Delta r_{\perp}$, where $P(r) \sim \rho$ is the radial 
  distribution function defined above (the number of other rods
  with a specified distance of closest approach per unit length
  of the test chain), $\Delta r_{\perp}$ is a typical magnitude for
  the transverse displacement of any rod of the chain, and $a
  \Delta r_{\perp}$ is comparable to the average area swept out 
  by one rod. The r.m.s. transverse displacement of an individual 
  bead or rod within a single time step of a discretized chain is 
  of order $\Delta r_{\perp} \sim \sqrt{ D_{b}\Delta t}$ of a 
  free bead, where $D_{b} = kT/(\zeta_{\perp} a)$ is the transverse 
  diffusivity of a single bead.  By combining these expressions, we 
  predict that the percentage of rejected moves scales as 
  \begin{equation}
     P_{reject} \sim A cL^{3} \sqrt{N \Delta t / \tauroddil}
     \label{Preject}
  \end{equation}
  where $\tauroddil = \zeta_{\perp}L^{3}/(72 kT)$ is the rotational 
  diffusion time of a rigid rod of length $L$ in dilute solution,
  and $A$ is a universal numerical prefactor.  This relation is 
  confirmed by the data shown in Figure \ref{fig:percentreject}, 
  where $P_{reject}$ is shown to be a linear function of $cL^{3} 
  \sqrt{ N \Delta t/\tauroddil}$ in systems with several values 
  of $cL^{3}$ and $N$. The dotted line in this figure yields a
  prefactor $A \simeq 0.078$.

\begin{figure}
    \begin{center}
      \gpfigc{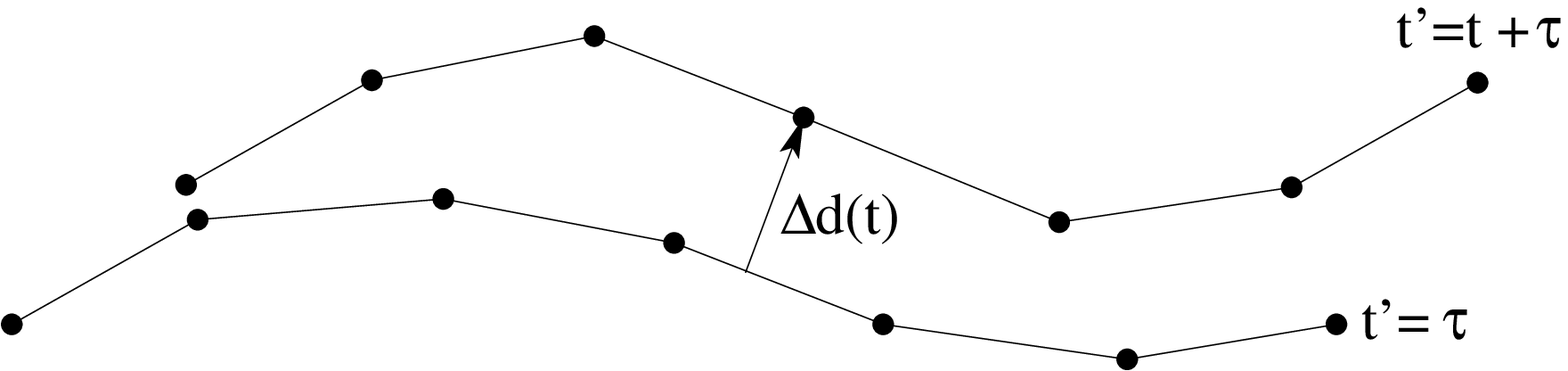}
      \caption[Schematic showing how to measure $ \msdtubedia $ ---
      the mean square displacement (MSD) of the middle bead relative
        to the tube]{Schematic diagram defining $\msdtubedia$ for a
        semiflexible beadrod chain with $N=6$}
      \label{fig:msdtubediaschematic}
    \end{center}
  \end{figure}

\begin{figure}
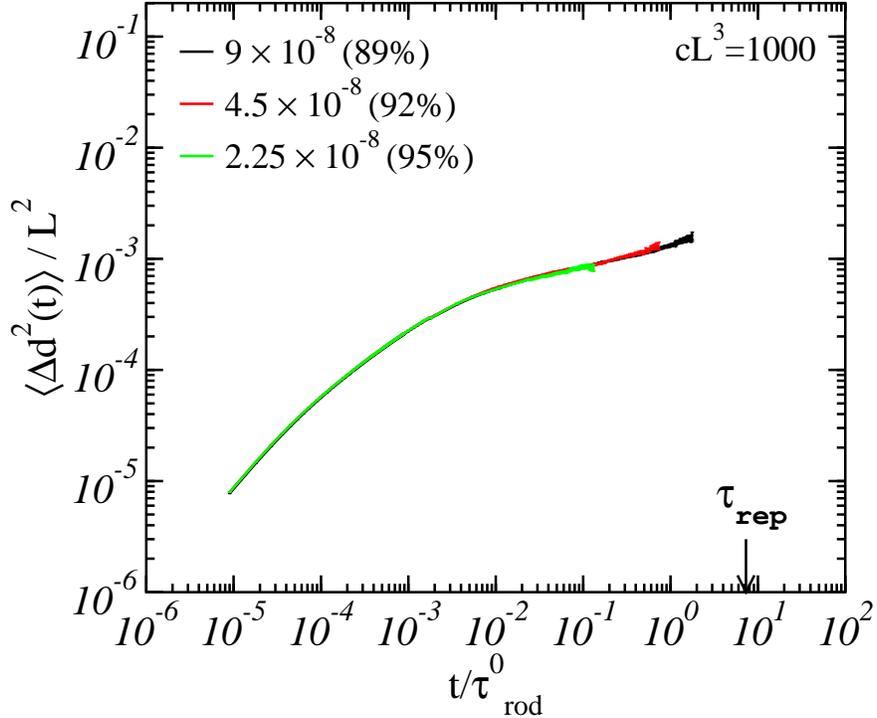

    \begin{center}
      \gpfigc{figs/tube-dia-cL3-1000-conv-deltat.eps}
      \caption[Convergence of $ \msdtubedia $ with respect to $ \Delta
      t $]{Convergence of $\msdtubedia$ with respect to integration
        timestep.  Legends indicate values of $ \Delta t / \tauroddil
        $.  Values in parenthesis represent the percentage of trial
        moves accepted.  $L_{p}=L$, $\zetapar=\zetaperp$, $N=20$ for
        all data shown above.}
      \label{fig:tube-dia-cL3-1000-conv-deltat}
    \end{center}
  \end{figure}

  Figure \ref{fig:tube-dia-cL3-1000-conv-deltat} shows the effect 
  of varying $\Delta t$ on a quantity $\msdtubedia$ that is
  approximately the mean square displacement of the middle bead 
  of a chain along directions transverse to the tube. The quantity 
  $\msdtubedia$ is defined to be the square of the distance of 
  closest approach between the position of the the middle bead 
  of a given polymer chain at a time $\tau+t$ and the contour 
  of the same chain at an earlier time $\tau$, as shown in 
  Figure \ref{fig:msdtubediaschematic}. In a tightly entangled 
  solution, the magnitude of the plateau in this quantity is a 
  measure of the width of the tube to which the polymer is confined, 
  and will be analyzed in detail elsewhere.  Clearly, the data 
  is well converged for all $t$ over the range of timestep values 
  shown here, for which the rejection ratio is of order 10 \%. 
  Similar results were observed for other dynamic properties. 

  Examination of this and other tests of the convergence with
  respect to time step led us to conclude that diffusion in 
  tightly entangled solutions could be accurately represented, 
  to within the statistical errors of our simulations, by 
  choosing a time step so as to accept at least 90 \% 
  of all trial moves. In highly entangled solutions, with 
  $cL^{3} > 250$ our choice of time step is controlled by 
  this criterion, rather than by the need to resolve the 
  fluctuations of individual chains.  Combining this criterion 
  with Eq. (\ref{Preject}) yields a simple prescription for 
  calculating $\Delta t$. 

\begin{figure}
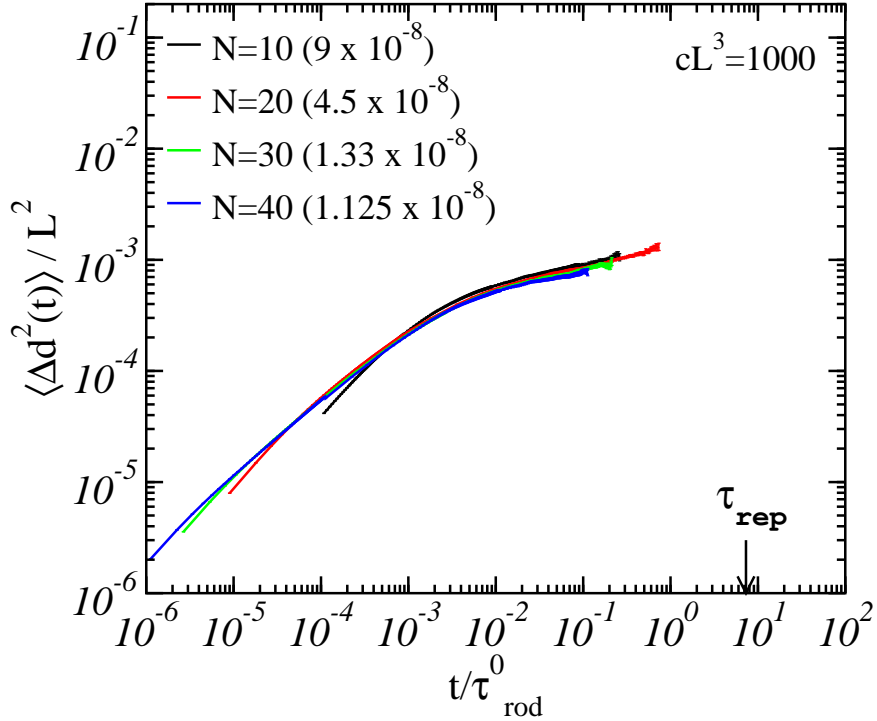

    \begin{center}
      \gpfigc{figs/tube-dia-cL3-1000-conv-nrods.eps}
      \caption[Convergence of $\msdtubedia$ with respect to chain
       discretization $N$]{Convergence of $\msdtubedia$ with respect to
       chain discretization $N$.  Legends indicate chain
       discretization $ N $.  Values in parenthesis represent $
       \Delta t / \tauroddil $.  $ \Delta t / \tauroddil $ has been
       chosen such that atleast $ 90\% $ of trial moves are accepted.
       $L_{p}=L$ and $\zetapar=\zetaperp$ for all data shown above.}
      \label{fig:tube-dia-cL3-1000-conv-nrods}
    \end{center}
  \end{figure}

  Figure \ref{fig:tube-dia-cL3-1000-conv-nrods} shows the 
  effect of changes in the number $N$ of rods per chain on 
  $\msdtubedia$ in solutions with $cL^{3} = 1000$. The 
  behavior of $\msdtubedia$ at very early times is, of 
  course, very sensitive to chain discretization. At longer 
  times, corresponding to the plateau in $\msdtubedia$, 
  the data is less strongly affected by changes in $N$. 

  \subsection{Computational requirements}
  \label{sec:time-mem-requirements}
  Table I compares simulation times and memory
  requirements from four different simulations of solutions of 
  semiflexible chains: (1) A simulation of isolated semiflexible 
  chains using an algorithm developed previously for simulating 
  dilute solutions, in which only the conformation of a single 
  chain is stored in memory, which uses the same time stepping
  algorithm subroutines as those used here. (2) A simulation of 
  6912 phantom chains using our code for simulating entangled
  solutions, but without checking for intersections or interactions.
  The primary difference between this and simulation 1 is that
  it requires the positions of all of the chains to be stored
  simultaneously in memory. (3) A simulation of an entangled 
  solution of 6912 uncrossable but infinitely thin semiflexible 
  threads, in which moves that cause chains to cross are rejected.
  (4) A simulation of corresponding solution of uncrossable chains 
  with repulsive excluded volume interaction, with a diameter 
  $d/L = 10^{-3}$, in which we must also calculate the change
  in interaction energy induced by moves that do not cause 
  chains to cross, in order to calculate the Metropolis 
  acceptance criterion. All simulations were performed on AMD 
  Athlon MP processors with a CPU speed of 1.67 GHz and a cache 
  size of 256 Kb, using a code that was implemented in Fortran 
  90. 

  From Table I, we see that the simulation of a solution of 
  phantom chains, in which all of the chain conformations are 
  stored but the chains do not interact, is about $8$ times 
  slower than the equivalent simulation of individual chains.
  This is presumably because the entire problem no longer 
  fits in the processors' cache. Once the cost of this change 
  in memory structure is paid, the addition of topological 
  checks slows down the simulation only by a factor of about 
  $1.5$. The addition of a small steric diameter causes an
  even smaller further increase.  The times per rod per time 
  step are quite comparable to the times of a few $\mu$s per 
  particle per time step required in MD simulations of bead
  spring polymers with repulsive Lennard Jones interactions 
  at liquid-like densities on the same processors. 

   \begin{table}
     \begin{center}
       \begin{tabular}{|l|c|c|c|}
         \hline
                     & Verlet      & Time           & Memory  \cr
         Simulation  & $r_{l}/L$   & $\mu$s /rod step & (MByte) \cr
         \hline
         1) Single   & $-$ & $0.64$ & $0.0054$ \cr
         \hline
         2) Phantom  & $-$ & $5.12$ & $81.7$ \cr
         \hline
         3) Entangled ($d=0$) & $0.0375$ & $7.61$ & $96.9$ \cr
         \hline
         4) Entangled ($d=10^{-3}L$) & $0.0375$ & $8.42$ & $96.9$ \cr
         \hline
       \end{tabular}
       \vspace{0.25in}
       \caption[Computational requirements] {Comparison of memory 
        and simulation times required for four different simulations 
        of solution of chains with $L_{p}=L$ and $N=40$, with a time 
        step $(\Delta t / \tauroddil) = 1.40625 \times 10^{-9}$. 
        Simulations (2-4) were of a system of 6912 molecules in a 
        periodic cell of volume chosen to give $cL^{3} = 4000$. }
     \end{center}
     \label{tab:efficiency}
   \end{table}

  \section{Theoretical Background}
  \label{sec:analysis}
  The algorithm presented in Sec. \ref{sec:simalgo} can be viewed
  either as a Brownian dynamics algorithm or as a form of Monte Carlo
  algorithm. The Brownian dynamics viewpoint is necessary to justify
  the use of the algorithm to predict dynamical properties. The Monte
  Carlo perspective is needed to explain the accuracy attained in
  results for equilibrium properties, and to justify our use of a
  Metropolis sampling scheme to treat the intermolecular potential.

  To highlight some of the conceptual issues that arise in the
  discussion of our simulations, we have found it useful to also
  consider stochastic algorithms for a related but simpler 
  problem of one-dimensional (1D) diffusion of a random walker 
  with a coordinate $x(t)>0$ near a hard wall at $x=0$. The 
  requirement that the walker cannot go through the wall is 
  analogous to the constraint prohibiting two chains from 
  cutting through each other. This constraint may be imposed 
  in the 1D diffusion problem by simply rejecting moves that 
  take the walker through the wall. 

  \subsection{Brownian dynamics}
  \label{sec:analysis-bd}
  When designing our algorithm for simulating infinitely thin
  uncrossable threads, we initially thought of it as a Brownian
  dynamics algorithm, in which the rejection of moves that lead to 
  chain crossings was viewed as being analogous to the rejection 
  scheme proposed above for simulating one-dimensional diffusion 
  near a reflecting boundary. 

  To clarify this point of view, and its limitations, we first 
  recall the theory underlying a conventional Brownian dynamics 
  algorithm. In this analysis, one considers a Markov jump process 
  for  some vector of coordinates $\vec{X} = [X_{1},\ldots,X_{n}]$. 
  in which the conditional probability distribution for any random 
  displacement $\vec{X} \rightarrow \vec{X} + \Delta  \vec{X}$ 
  depends upon an adjustable time step  $\Delta t$. The relevant 
  set of coordinates $\vec{X}$ is the set of all bead positions 
  in a simulation of a polymer solution  or the coordinate $x$ of 
  the random walker in the 1D diffusion problem. The probability
  distribution $P(\vec{X},t)$ generated by this discrete process 
  may be shown to converge in the limit $\Delta t \rightarrow 0$ 
  to the  solution of a corresponding Fokker-Planck equation
  \begin{equation}
    \frac{\partial P}{\partial t} = 
    - \frac{\partial (V_{i}P)}{\partial X_{i}}
    + \frac{\partial (D_{ij} P)}{\partial X_{i}\partial X_{j}}
  \end{equation}
  if the first and second moments of the random displacement 
  $\Delta X_{i}$ from a state $\vec{X}$ obey the conditions
  \begin{eqnarray}
     V_{i}(\vec{X}) & = & \lim_{\Delta t \rightarrow 0}
     \frac{\langle \Delta X_{i} \rangle}{\Delta t}
     \nonumber \\
     D_{ij}(\vec{X}) & = & \lim_{\Delta t \rightarrow 0}
     \frac{\langle \Delta X_{i} \Delta X_{j}\rangle}{2 \Delta t}
     \label{BD-moments}
  \end{eqnarray}
  where $V_{i}(\vec{X}) $ is a drift velocity vector and
  $D_{ij}(\vec{X})$ is a diffusivity tensor.  The BD algorithm 
  for non-interacting polymers used in our simulations was
  designed so as to satisfy these conditions for the drift 
  velocity and diffusivity appropriate to a wormlike chain.
  \cite{fixman1978,hinch1994,Morse2004}.

  In our simulations of a system of $M$ polymers, the relevant
  underlying Markov step is the motion of a single randomly 
  chosen chain. The actual time elapsed during the motion of 
  a single chain must be understood to be smaller by a factor
  of $M$ than the timestep  of $\Delta t$ used in the underlying 
  single-chain algorithm.  This trivial rescaling of $\Delta t$ 
  is needed to compensate for a  corresponding reduction of the 
  values of the moments  $\langle \Delta X_{i} \rangle$ and
  $\langle \Delta X_{i}  \Delta X_{j} \rangle$ in Eq.
  (\ref{BD-moments}) for coordinates $X_{i}$ and $X_{j}$, which
  correspond to Cartesian components  of the positions of beads 
  on the same chain: Both of these  moments are reduced by a 
  factor of $1/M$ relative to the values obtained in a 
  single-chain simulation because there is only a probability 
  $1/M$ that any particular chain will move during a time step.
 
  The derivation of the analysis outlined above relies upon the
  assumption that both $\vec{V_{i}}(\vec{X})$ and $D_{ij}(\vec{X})$
  are smooth functions of $\vec{X}$. This condition is not satisfied
  by diffusion near a reflecting boundary, in which the underlying
  diffusion equation must either be taken to contain values of 
  $V_{i}(\vec{X})$ and $D_{ij}(\vec{X})$ that jump discontinuously 
  to zero at a boundary, or must be explicitly supplemented in the 
  Fokker-Planck equation by a reflecting boundary condition.  Several
  authors have considered how to correctly implement a  reflecting
  boundary condition in a stochastic simulation, and proposed methods
  for minimizing the error arising from the presence of such a 
  boundary\cite{Schulten1983,Ottinger1989,Peters2002}. The correct
  probability distribution function $P(\vec{X},t)$  can actually be
  obtained in the limit $\Delta t \rightarrow 0$ from a variety
  algorithms in which the random walker is prevented from penetrating
  the wall. For example, for a 1D walker confined by a wall to $x>0$, 
  the same probability distribution is obtained in the limit $\Delta 
  t  \rightarrow 0$ either by rejecting moves that yield a trial 
  position $x<0$, or by replacing trial values of  $x < 0$ by $-x$.  
  The choice of an algorithm for treating jumps near a reflecting 
  boundary does, however, affect the magnitude of the systematic
  time-discretization error produced by a stochastic simulation 
  for  $\Delta t \neq 0$.  In the simple 1D diffusion problem, in 
  which  each particle jumps a distance of order $\Delta x \sim 
  \sqrt{D \Delta t}$  per time step, most ``plausible'' rules for 
  treating jumps near a reflecting boundary introduce a large, 
  ${\cal O}(1)$ error in $P(x)$ over a region of size 
  ${\cal O}(\Delta x)$ of the boundary, and produce smaller 
  ${\cal O}(\sqrt{\Delta t})$ errors far from the wall.

  In light of this understanding, we thus initially expected our 
  algorithm to yield a radial distribution function $P(r)$ for 
  the distance of closest approach between pairs of uncrossable 
  rods that deviates significantly from theoretical predictions 
  for  inter-rod distances $r$ of order the typical displacement 
  $\Delta x$ of a single bead over one time step, which is given 
  by $\Delta x \sim \sqrt{k_{B}T \Delta t/(a\zetaperp)}$.  We 
  were surprised to find instead that there was no measurable 
  deviation of $P(r)$  from the theoretical predictions, even at 
  extremely small values  of $r$, as shown in Figure 
  \ref{fig:raddist}.  Our understanding  of this result is based 
  upon an interpretation of the algorithm as a form of Monte Carlo sampling.

  \subsection{Monte Carlo}
  \label{sec:analysis-mc}
  Monte Carlo simulation is a method of efficiently sampling a 
  desired equilibrium probability distribution $P_{eq}(\vec{X})$ 
  for some set of coordinates $\vec{X}$. A Monte-Carlo simulation 
  is based on a Markov process in which the conditional probability 
  $T( \vec{X}_{i} \rightarrow \vec{X}_{f})$ of a transition from a 
  state $\vec{X}_{i}$ to a state $\vec{X}_{f}$ is related to the
  probability of the reverse transition by a detailed balance
  condition \citep{binder}
  \begin{equation}
    P_{eq}(\vec{X}_{i}) T( \vec{X}_{i} \rightarrow \vec{X}_{f}) =
    P_{eq}(\vec{X}_{f}) T( \vec{X}_{f} \rightarrow \vec{X}_{i})
    \quad.
  \end{equation}
  If the Monte-Carlo jump process involves both the generation of
  a trial move, and a decision regarding whether to accept or reject 
  the move, then
  \begin{equation}
    \label{eq:transition-matrix}
    T(\vec{X}_{i} \rightarrow \vec{X}_{f}) = 
    G(\vec{X}_{i} \rightarrow \vec{X}_{f}) 
    A(\vec{X}_{i} \rightarrow \vec{X}_{f}) 
  \end{equation}
  where $G(\vec{X}_{i} \rightarrow \vec{X}_{f})$ is the probability
  of generating a particular trial move and $A(\vec{X}_{i}  \rightarrow
  \vec{X}_{f})$ is an acceptance probability. 
  
  \subsubsection{Uncrossable Threads}
  We first consider the algorithm used to simulate infinitely thin
  uncrossable polymers. The only potential energy in the problem is 
  the intramolecular bending energy $U_{0}$ of all of the chains.
  In our algorithm, trial moves are generated by a  Brownian
  dynamics algorithm that has been designed so as to  generate an
  equilibrium distribution for this potential. We thus consider a 
  class of algorithms in which a BD algorithm is used to generate 
  trial moves, and in which the BD algorithm is assumed (for the
  moment) to approximately satisfy detailed balance in the limit 
  $\Delta t \rightarrow 0$. That is, we assume for the moment that
  \begin{equation}
    P_{0}(\vec{X}_{i}) G( \vec{X}_{i} \rightarrow \vec{X}_{f}) = 
    P_{0}(\vec{X}_{f}) 
    G( \vec{X}_{f} \rightarrow \vec{X}_{i}) 
    \label{eq:BD-DetailedBalance}
  \end{equation}
  in the limit $\Delta t \rightarrow 0$ in which the BD algorithm 
  is designed to yield the equilibrium distribution $P_{0}(\vec{X})$.
  Here, $P_{0}(\vec{X})\propto e^{-U_{0}(\vec{X})}$ is the 
  equilibrium distribution for a solution of non-interacting 
  wormlike chains, which is also the correct equilibrium 
  distribution for a solution of infinitely thin but uncrossable 
  chains with no interaction potential. 

  The prohibition on chain crossing is enforced by simply rejecting 
  all moves that cause one chain to cut through another ($A=0$) and
  accepting all others ($A=1$).  If a transition $\vec{X}_{i} 
  \rightarrow \vec{X}_{f}$ involving the motion of the beads of 
  one chain causes that chain to cut through another, then the 
  hypothetical reverse move $\vec{X}_{f} \rightarrow \vec{X}_{i}$,
  which would involve moving the same chain from its final to 
  initial state, would also cause a chain crossing, and so would 
  also be rejected. If the matrix $G$ satisfies detailed balance, 
  but certain types of moves are prohibited, then the overall 
  transition matrix $T=GA$ will thus also satisfy detailed 
  balance, as long as the reverse of any prohibited move is also 
  prohibited, as is the case here. An algorithm in which a BD 
  algorithm is supplemented by such a rejection scheme will thus 
  satisfy detailed balance to the same extent as the underlying 
  BD algorithm. If the BD simulation can be shown to satisfy (or
  nearly satisfy) detailed balance, this would explain the 
  otherwise surprising accuracy of our results for $P(r)$.
  
  \subsubsection{Intermolecular Interactions}
  Our algorithm for simulating a system of chains with a 
  short range repulsive intermolecular repulsion is an extension 
  of the above idea.  We express the total potential energy 
  of the system as a sum
  \begin{equation}
    U = U_{0} + U_{int}
  \end{equation}
  where $U_{0}$ is a relatively ``soft'' intramolecular bending 
  energy and $U_{int}$ is a ``hard'' intermolecular interaction 
  energy. The equilibrium distribution for the interacting system 
  is of the form
  \begin{equation}
    P_{eq}(\vec{X}) = P_{0}(\vec{X})e^{-U_{int}(\vec{X})/k_{B}T }
  \end{equation}
  where $P_{0}(\vec{X})$ is the equilibrium distribution for the
  non-interacting gas, which is governed by the bending energy. 
  If moves are generated with a BD algorithm that satisfies
  detailed balance condition \eqref{eq:BD-DetailedBalance} for
  non-interacting polymers, then the corresponding condition 
  for the interacting solution can be satisfied by requiring 
  the acceptance probability $A$ to satisfy
  \begin{equation}
    \frac{A(\vec{X}_{i} \rightarrow \vec{X}_{f}) }
    { A(\vec{X}_{f} \rightarrow \vec{X}_{i}) }
    = e^{ -\Delta U_{int}/k_B T }
  \end{equation}
  where $\Delta U_{int} \equiv U_{int}(\vec{X}_{f})-U_{int}(\vec{X}_{i})$
  is the change in interaction energy.  We accomplish this by 
  supplementing our prohibition on chain crossing by the usual
  Metropolis scheme of taking $A(\vec{X}_{i} \rightarrow \vec{X}_{f} ) 
  = 1 $ for $\Delta U_{int} < 0$ and $A( \vec{X}_{i} \rightarrow 
  \vec{X}_{f} ) = e^{-\Delta U_{int}/k_{B}T} $ for $\Delta U_{int} 
  > 0$ for moves that do not cause chains to cross. The same 
  reasoning underlies our slithering snake algorithm for 
  equilibrating such systems, in which the trial moves are
  slithering snake moves that are designed to satisfy detailed
  balance for non-interacting chains. 
 
  Our use of a Brownian dynamics algorithm to generate trial 
  moves and a Metropolis acceptance scheme based on the interaction 
  energy is useful in situations in which we wish to resolve the 
  dynamics of chain bending, and $\Delta t$ can be made small 
  enough to do so, but in which we do not need to resolve the 
  dynamics of intra-chain collisions, and in which the range of 
  repulsion between chains is so small that dynamically 
  resolving collisions would require us to use a much smaller 
  time step. Analogous hybrid algorithms are potentially useful 
  in any problem in which one wishes to simulate diffusion in
  a comparatively soft potential in the presence of an
  additional steeply repulsive potential.
  
  \subsubsection{Do BD simulations satisfy detailed balance?}
  
  In appendix \ref{app:1D-diffusion}, we try to answer  the question 
  of whether a valid Brownian dynamics algorithm satisfies detailed 
  balance by considering a simplified  model of one dimensional (1D) 
  diffusion under the influence of a soft potential $U_{0}(X)$. There 
  we consider both an algorithm in which the  random force is chosen 
  from a uniform distribution and  an algorithm in which the random 
  force is chosen from  a Gaussian distribution. We find that, for 
  this model, both algorithms satisfy detailed balance for simple 
  diffusion in a constant potential, with no drift. In  problems with 
  nonzero deterministic drift forces, it is straightforward to show 
  that the BD algorithm with normally distributed random forces 
  satisfies detailed balance for any linear potential, and should 
  thus satisfy detailed balance in the limit of $\Delta t \rightarrow 0$ 
  for any locally linear potential. It is equally easy to show that
  an algorithm with a uniformly distributed random force in addition
  to a deterministic force does not exactly satisfy detailed balance.
  It is argued, however, that the small violation of detailed balance 
  found with uniformly distributed noise should have a negligible 
  effect when  the step size is taken to be small enough to resolve 
  variations in the soft potential. We thus conclude that, with 
  either type of BD algorithm, the effect of any violation of detailed 
  balance should vanish in the limit $\Delta t \rightarrow 0$ of 
  interest.
  
\begin{figure}
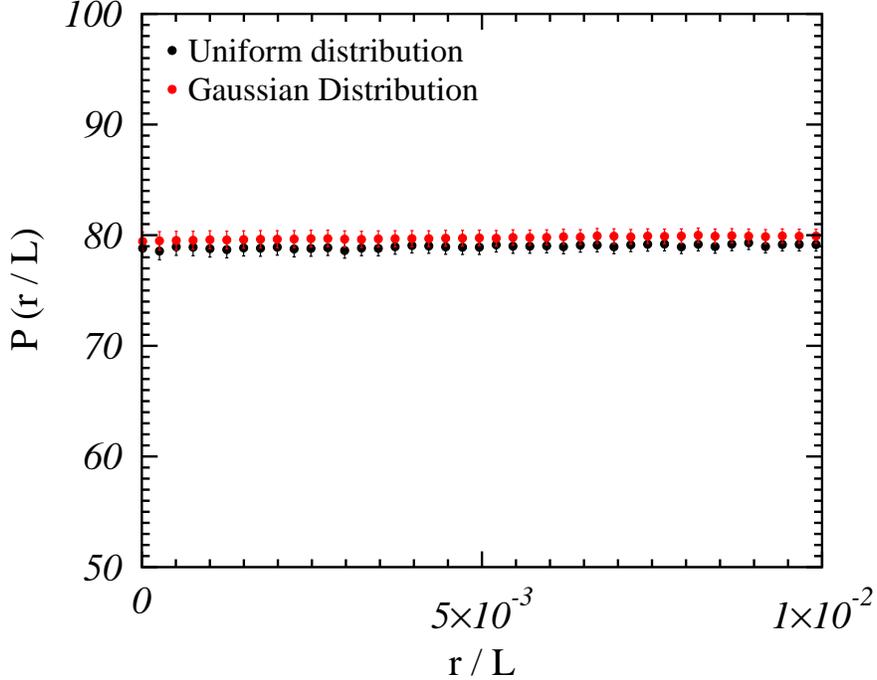

    \begin{center}
      \gpfigc{figs/radial-dist-conv-noise-2.eps}
      \caption{Effect of noise distribution on P(r). $ cL^{3}=1000 $,
      $N=20$, $ L_{p}=L$, $\zetapar=\zetaperp$, $ \Delta t /
      \tauroddil = 4.5 \times 10^{-8} $.}
      \label{fig:rad-dist-conv-noise}
    \end{center}
  \end{figure}

  Our conclusions about the simple 1D diffusion problem are
  consistent with our results for simulations of entangled polymers,
  in which we find that the combination of single-chain BD simulation
  with an appropriate  rejection criterion leads to results that agree
  to within stochastic error with theoretical  predictions for the
  radial distribution and other equilibrium properties. In most of 
  our simulations, we have used an unprojected random force $\etav'$ 
  whose Cartesian components are chosen from a uniform distribution.  
  An extensive series of simulations was carried out with this choice 
  before we understood its potential theoretical significance. In 
  order to check whether our use of uniform distribution affected 
  our results, we also ran a smaller number of simulations with a 
  Gaussian distribution for $\etav'$.
  Figures \ref{fig:rad-dist-conv-noise} and 
  \ref{fig:tube-dia-cL3-1000-conv-noise} compare results for 
  $P(r)$ and $\msdtubedia$ obtained from simulations of infinitely
  thin threads using these two different noise distributions. 
  For the value of $\Delta t$ used in our simulations, the results
  agree to within statistical error. Because the mathematical
  argument required to show that detailed balance is obtained in 
  the limit $\Delta t \rightarrow 0$ is relatively straightforward 
  only for BD algorithms with Gaussian-distributed random forces, 
  however, we recommend that any future applications of a hybrid 
  BD-MC algorithm analogous to that used here use Gaussian 
  distributed random forces, if only to simplify any subsequent 
  discussions of the underlying theory. 

  \begin{figure}
    \begin{center}
      \gpfigc{figs/tube-dia-cL3-1000-conv-noise.eps}
      \caption{Effect of noise distribution on $\msdtubedia$.
      $cL^{3}=1000$, $N=20$, $L_{p}=L$, $\zetapar=\zetaperp$,
      $\Delta t / \tauroddil = 4.5 \times 10^{-8} $.}
      \label{fig:tube-dia-cL3-1000-conv-noise}
    \end{center}
  \end{figure}

  \section{Conclusions}
  \label{sec:conclusions}

  We have presented a hybrid algorithm for simulating highly
  entangled solutions of very thin uncrossable wormlike 
  filaments.  The algorithm uses a single-chain Brownian dynamics 
  (BD) algorithm for generating trial moves.  This is supplemented 
  with a Monte Carlo-like scheme where any trial move that causes 
  two chains to cross or that leads to excluded volume overlaps 
  are rejected. Efficient methods were developed for checking for 
  chain crossings and excluded volume violations.  Because the
  algorithm is based on a Brownian dynamics time-stepping 
  algorithm, it can be used to simulate dynamical, as well 
  as equilibrium properties. The algorithm is expected to be 
  much more efficient than Brownian dynamics simulation of a 
  bead-spring model of nearly tangent beads in the limit of 
  highly entangled solutions of extremely thin chains. Comparison 
  with the efficiency of a molecular dynamics simulation of a 
  bead spring model is complicated by the difference between the 
  diffusive dynamics of our algorithm and the ballistic motion 
  obtained in molecular dynamics of a semi-dilute solution in 
  the absence of explicit solvent, which would lead to efficient 
  sampling of configuration space but a loss of dynamical realism. 
  We have used the algorithm discussed here to study dynamics 
  and stress relaxation in highly entangled solutions
  \cite{Ramanathan_thesis}, which will be discussed elsewhere. 
  

  \newcommand{\appsection}[1]{\let\oldthesection\thesection
    \renewcommand{\thesection}{\oldthesection}
    \section{#1}\let\thesection\oldthesection}

  \appendix

    
  \appsection{Radial distribution function for rods}
  \label{app:radial-dist-fn}
  \setcounter{equation}{0}
  Consider a test rod oriented along the z-axis.
  Let $\rv$ be the closest approach
  vector and $\nv$ be the unit vector normal to the plane defined 
  by the test rod and the closest approach vector.  Choose an
  infinitesimal area on this plane, $ d\Av = d\rv \times d\zv $.
  Since $ d\rv $ is perpendicular to $ d \zv $ and $ \nv $ is
  perpendicular to this area unit, one can write $ d\Av = dr dz \nv $.
  Using geometry, one can show that the number of rods, with
  orientations lying between $ \uv $ and $ \uv + d\uv $, that pierce
  this infinitesimal area is given by
  \begin{equation}
    \label{eq:delta-no-rods-piercing-plane}
    dN_{\uv} = \rho \; d\Av \cdot \uv \; (2) \; \frac{d^{2} \uv}{4 \pi}.
  \end{equation}
  where $ \rho $, as defined earlier, represents the contour length
  density.  $ d^{2} \uv $ represents the solid angle subtended by $
  \uv $ and $ \uv + d \uv $ and is given by $ d^{2} \uv = \sin \theta
  d \theta d \phi $, where $ \theta $ represents the angle between $
  \uv $ and the z-axis and $ \phi $ represents the azimuthal angle.
  The factor $ 4 \pi $ in the denominator is obtained by integrating
  with respect to $ d^{2} \uv $ over the surface of a sphere.  The
  factor of $2$ in the numerator arises because of the need to count
  for rods with both orientations, $ \uv $ and $ - \uv $.
  Substituting for $ d^{2} \uv $ and $ d\Av $ from above while noting
  that $ \nv \cdot \uv = \cos (90^{o}-\theta) = \sin \theta $ and
  integrating equation \eqref{eq:delta-no-rods-piercing-plane} over $
  d\theta $, $ d\phi $, we obtain the radial distribution function
  given in Eq. (\ref{eq:raddistfn}). 

  \appsection{Simulation of one-dimensional diffusion}
  \label{app:1D-diffusion}
  \setcounter{equation}{0}
  In this appendix, we consider whether a very simple Brownian
  dynamics algorithm satisfies detailed balance when applied to 
  one dimensional (1D) diffusion. We consider the use of a
  hybrid BD-MC algorithm analogous to that discussed in the 
  body of the paper to describe the diffusion of a particle
  with a position $X(t)$ which is subjected to both a 
  slowly-varying (or ``soft'') potential $U_{0}(X)$ and an 
  arbitrary steep repulsive (or ``hard'') potential $U_{1}(X)$ 
  that represents the walls of a confining 1D box. To represent
  a box of length $L$ with hard walls, we might take $U_{1}(X)$ 
  to vanish for $-L/2 < X < L/2$, and to diverge outside this 
  domain.  We consider an algorithm in which trial moves are 
  generated by a Brownian dynamics algorithm that is obtained 
  by discretizing the Langevin equation 
  \begin{equation}
    \label{eq:particle-langevin-eqn}
    \zeta dX = 
   \left[ F_{0}(X) + \eta (t) \right]dt
  \end{equation}
  where $\zeta$ is a friction coefficient, $X$ is the particle 
  position, $\eta (t)$ is a random force, and $F_{0}(X) = -
  d U(X)/dX$ is a force arising from the soft potential. Each 
  time step, we generate a trial step $X_{i} \rightarrow X_{f}$
  of magnitude
  \begin{equation}
    \label{eq:particle-DX}
    \zeta \Delta X = \left [ F_{0}(X_{i} ) + \eta \right ] 
    \Delta t 
  \end{equation}
  where $\Delta X = X_{f} - X_{i}$, and where $\eta$ is chosen 
  from a distribution $P(\eta)$ with moments 
  $\langle \eta \rangle = 0 $ and $\langle \eta^{2} \rangle 
  = 2 k_{B}T \zeta / \Delta t $. Here, we consider whether 
  the resulting jump probability satisfies the detailed balance 
  condition
  $P_{0}(X_{i})G(X_{i} \rightarrow X_{f}) = P_{0}
  (X_{f})G(X_{f} \rightarrow X_{i})$,
  where $P_{0}(X) \propto e^{-U_{0}(X)/k_{B}T}$.  If, in a given 
  step, we generate a trial move $X_{i} \rightarrow X_{f}$ using 
  a force $\eta^{+}$, we must consider a hypothetical reverse 
  move in which the random force $\eta^{-}$ must satisfy
  \begin{equation}
     \eta^{+} + F_{0}(X_{i}) =  -[ \eta^{-} + F_{0}(X_{f})]
  \end{equation}
  in order to generate a displacement $-\Delta X$. 

  If the probability distribution $P(\eta)$ for the random force 
  $\eta$ is a Gaussian, 
  \begin{equation}
     P(\eta) \propto \exp\left ( - \eta^{2}/(4 k_{B} T\zeta)\right )
     \; ,
  \end{equation}
  it is straightforward to show that
  \begin{equation}
    \frac{ G(X_{i} \rightarrow X_{f}) }
         { G(X_{f} \rightarrow X_{i}) } 
    = e^{- \bar{F} [ \eta + \bar{F} ] \Delta t/(\zeta k_{B} T) }
    \;
  \end{equation}
  where $\bar{F} \equiv [ F_{0}(X_{i}) + F_{0}(X_{f})]/2$.  
  To lowest order in $\Delta X$, we may approximate the 
  change in energy $\Delta U \equiv U(X_{F})-U(X_{i})$ by 
  $\Delta U \simeq \bar{F}\delta X$ and approximate
  $\Delta X \simeq [\eta + \bar{F}]\Delta t$  to show that
  \begin{equation}
  \frac{ G(X_{i} \rightarrow X_{f}) }
       { G(X_{f} \rightarrow X_{i}) } 
       \simeq e^{-\Delta U_{0} /k_{B}T} 
       \quad.
  \end{equation}
  The algorithm thus approximately satisfies the detailed 
  balance condition in the limit $\Delta t \rightarrow 0$ 
  of interest, and can be shown to exactly satisfy detailed
  balance in the special case of a constant force $F_{0}(X)$.

  We next consider a uniformly distributed random force, for 
  which $P(\eta) = 1/(2E)$ over a range $-E < \eta < E$, where 
  $E = \sqrt{6\zeta k_{B}T/\Delta t}$.  This algorithm clearly
  does not exactly satisfy detailed balance, since a step for which 
  the random force $\eta^{+}$ that generates the forward trial move 
  is near one of the boundaries $\pm E$ can sometimes be reversed 
  only by a random force $\eta^{-}$ that lies outside the allowed 
  range $[-E,E]$. An algorithm in which the reverse of an allowed
  step is sometimes not allowed clearly cannot exactly satisfy 
  detailed balance.  In the limit of small force $F_{0}$, however, 
  this violation becomes small, insofar as it only affects a small 
  fraction of all jumps for which $\eta$ falls within a narrow 
  range of values near its maximum or minimum allowed value of
  $\pm E$. In the special case of a vanishing force, or constant 
  potential $U_{0}(X)$, detailed balance is recovered, for any
  distribution with the symmetry $P(\eta) = P(-\eta)$, including 
  a uniform distribution. 
    
  In order to assess the importance of this violation of detailed 
  balance for an algorithm with uniformly distributed noise and a
  spatially varying potential $U_{0}$, we must consider the
  implications of our original assumption that $U_{0}(X) $ is 
  a ``soft'' potential and $U_{1} (x)$ is ``hard''.  The effect of 
  a deterministic force $F$ becomes comparable to or greater than 
  that of diffusion only when applied over a characteristic distance 
  of order $k_{B}T/F$ or greater, for which the corresponding potential 
  energy exceeds $k_{B}T$, or over a corresponding time scale of 
  order $k_{B}T\zeta/F^{2}$, for which the displacement $F t/\zeta$ 
  caused by drift exceeds the root-mean-squared displacement 
  $\sqrt{D t}$ caused by diffusion. If the range of the hard potential 
  $U_{1}$ is much less than the characteristic length scale $k_{B}T/F$, 
  then we expect the existence of a nonzero force $F$ to have 
  negligible effect upon the structure of the probability 
  distribution within the narrow boundary layer in which the hard 
  repulsive potential is important, other than to change an 
  overall prefactor that is sensitive to the nature of the far 
  field.  Within this boundary layer, the form of the probability 
  distribution is instead determined by a balance between the 
  repulsive potential $U_{1}$ and diffusion. Since the form of 
  the equilibrium distribution produced by this balance can be 
  correctly simulated by an algorithm with no drift, which does 
  satisfy detailed balance, we expect the addition of a small 
  drift term (and a correspondingly small violation of detailed 
  balance) to have little effect upon the form of the probability 
  distribution near the boundary, as long as the range of the 
  ``hard'' potential $U_{1}$ is much less than the characteristic 
  length scale $k_{B}T/F$ of the ``soft'' potential. 
 
  For completeness we may also consider the case in which the 
  range of the repulsive potential $U_{1}$ is actually comparable 
  to that the characteristic length scale of the soft potential 
  $U_{0}$. Since we rely on a BD algorithm to resolve the effect 
  of potential $U_{0}$, we assume that we will choose a step size 
  $\Delta t$ such that the corresponding spatial step $\Delta X 
  \simeq \sqrt{D \Delta t}$ is much less than any length 
  characteristic of potential $U_{0}$. If the characteristic length 
  scales of $U_{0}$ and $U_{1}$ are comparable, however, then we 
  will automatically have chosen the step size small enough so 
  that potential $U_{1}$ changes by much less than $k_{B}T$ per 
  step. It can be shown in this case, by considering the effect 
  of the acceptance criterion in the limit of a soft potential 
  $U_{1}$, that the overall transition matrix $T=GA$ for the 
  resulting algorithm satisfies the conditions on the first and 
  second moments necessary to obtain a valid Brownian dynamics 
  algorithm for the total potential $U_{0}+U_{1}$. In this case, 
  the hybrid algorithm can thus be justified as a valid form of 
  Brownian dynamics simulation, and the violation of detailed 
  balance becomes irrelevant.



  

   
\end{document}